\documentclass[prl,aps,twocolumn]{revtex4-1}

\usepackage{times}
\usepackage{graphicx}
\usepackage{float}
\usepackage{latexsym,amsmath,amssymb,bm,euscript}
\usepackage{color}
\usepackage{subfigure}
\usepackage{epstopdf}
\usepackage[colorlinks=true,linkcolor=blue,citecolor=blue]{hyperref}
\usepackage{type1cm}

\usepackage{ulem}
\usepackage{cancel}

\usepackage{appendix}

\newcommand{\beginsupplement}{
\setcounter{table}{0}
\renewcommand{\thetable}{S\arabic{table}}
\setcounter{figure}{0}
\renewcommand{\thefigure}{S\arabic{figure}}
}

\begin{document}

\title{Topological Interface between Pfaffian and anti-Pfaffian Order in $\nu=5/2$  Quantum Hall Effect}

\author{W. Zhu$^1$, D. N. Sheng$^2$, and Kun Yang$^3$}
\affiliation{$^1$ Westlake Institute of Advanced Study, Westlake University, Hangzhou, 310024, China}
\affiliation{$^2$Department of Physics and Astronomy, California State University, Northridge, CA 91330, USA}
\affiliation{$^3$National High Magnetic Field Laboratory and Physics Department, Florida State University, Tallahassee, FL 32306, USA}

\begin{abstract}
Recent thermal Hall experiment pumped new energy into the problem of $\nu=5/2$  quantum Hall effect,
which motivated novel interpretations based on formation of mesoscopic puddles made of Pfaffian and anti-Pfaffian topological orders.
Here, we study an interface between the Pfaffian and anti-Pfaffian states, which may play crucial roles in thermal transport,
by means of state-of-the-art density-matrix renormalization group simulations on the cylinder geometry.
We provide compelling evidences that indicate the edge modes of the Pfaffian and anti-Pfaffian state
strongly hybridize with each other around the interface.
Moreover, we demonstrate an intrinsic electric dipole moment emerges
at the interface, similar to the ``p-n" junction sandwiched between N-type and P-type semiconductor.
Importantly, we elucidate the topological origin of this dipole moment,
whose formation is to counterbalance the mismatch of guiding-center Hall viscosity of bulk Pfaffian and anti-Pfaffian state.
\end{abstract}

\date{\today}


\maketitle


The $\nu = 5/2$ fractional quantum Hall (FQH)  effect in the second Landau level
has sparked much interest in condensed matter for decades 
\cite{Willett1989,Pan1999,Pan2008,Choi2008,Dolev2008,Radu2008,Bid2010,Willett2013,Stern2010,Tiemann2008,Baer2014,Yacoby2011},
mainly due to its likely non-Abelian nature and potential application in topological quantum computation \cite{Kitaev2003,Nayak}.
The leading theoretical candidate is the non-Abelian Pfaffian (Pf) state \cite{Moore1991,Greiter1991},
a fully polarized chiral p-wave state of composite fermions \cite{Read2000},
as supported by numerical studies \cite{Morf1998,Rezayi2000,Wan2006,PhysRevB.80.235330,Moller2008,Peterson2008,HaoWang2009,Wojs2010,Morf2010,Feiguin2008,Storni2011,Pakrouski2015,WZhu2016}.
Besides, its particle-hole conjugate partner,
known as the anti-Pfaffian (APf) state \cite{Levin2007,SSLee2007}, is an equally valid candidate, 
which may actually be more viable under realistic experimental conditions \cite{Zalatel2015,Rezayi2017}.
Breaking of particle-hole symmetry, either spontaneously or explicitly, is crucial for the emergence of the Pf or APf state.
Recently, to interpret the observation of half-integer thermal Hall conductance that is consistent with particle-hole symmetry \cite{Banerjee2018},
the particle-hole preserved Pf state was proposed \cite{Feldman2016,XChen2014}.
Alternatively, the experimental observation could be simply explained by lack of thermal equilibration at the edge\cite{Simon2018b}
(a scenario currently under debate \cite{Feldman2018,Simon2018a}), or more significantly,
by the presence of random domains made of the Pf and APf states \cite{Mross2018,Chong2018,Lian2018},
similar to an earlier proposal of spontaneously formed Pf and APf strips \cite{Wan2016}.
The latter makes understanding of Pf-APf domain walls an urgent priority.

Generally speaking, the topologically protected edge states directly reflect bulk topological order via the bulk-edge correspondence \cite{Wen_book,Wen1993,Haldane2008,XLQi2012}, rendering edge the preferred window to peek into the fascinating bulk physics in topological states of matter \cite{DasSarma2005,Ady2006,Bonderson2006,Fendley2006}.
In the particular case of non-Abelian Pf-type states, this correspondence leads to the presence of neutral
Majorana fermion modes at the edge \cite{Milovanovic1996} (i.e. the interface separating the bulk from vacuum) responsible for the half-integer quantized thermal Hall conductance\cite{Banerjee2018}.
Relatively speaking less attention is drawn to the interface between two distinct topological states
\cite{Nancy1998,Kapustin2011,Kitaev2012,Barkeshli2013,Levin2013,YMLu2014,Cano2015,PhysRevB.96.241305,Santos2018,Regnault2019a,Regnault2019b,Jaworowski2019},
especially for those separating two non-Abelian orders \cite{Grosfeld2009,Slingerland2009,Barkeshli2015,Wan2016}.
Existing theoretical attempts mostly rely on the effective field theories, where
novel phenomena may emerge through the coupling between the two edges that meet at the interface.
While such phenomenological theory is good at
obtaining a qualitative understanding of the possible phases, 
many open questions remain and call for
quantitative study by unbiased numerical approaches \cite{Regnault2019a,Regnault2019b}.
For example,
it is extremely difficult for effective theories to determine which interface state is energetically favored by the microscopic interactions,
as well as non-universal aspects like edge reconstruction\cite{PhysRevB.49.8227,PhysRevLett.88.056802,PhysRevB.68.125307} 
which in principle could also happen at the interface \cite{PhysRevB.96.241305}. 
Numerical simulation is expected, in a
quantitative and unbiased way, to overcome these challenges faced by effective field theories. 
It is therefore highly desirable and urgent to develop an advanced numerical scheme,
to address  some pressing problems like the Pf-APf interface.

In this paper, we construct an interface between the Pf and APf state,
based on which we investigate the underlying physics of Pf-APf domain wall in the FQH effect at the filling factor $\nu=5/2$.
Our approach is based on a  design of cylinder geometry, by utilizing the density-matrix renormalization group (DMRG) algorithm.
We establish that the edge modes of the Pf and APf state strongly hybridize near the interface,
indicating that counter-propagating charge modes are fully gapped out.
Moreover, we identify the appearance of charge inhomogeneity around the interface, 
which yields a robust electric dipole moment. 
Crucially, we identify  the mismatch of Hall viscosity between the Pf and APf topological
orders as the driving force behind this dipole moment,
thus revealing the topological content of the Pf-APf interface, 
whose possible experimental consequences will be discussed.

\textit{Model and Method.---}
We consider interacting electrons in the presence of a perpendicular magnetic field on the cylinder geometry.
In the Landau gauge $\mathbf{A}=(0,Bx)$, the single-particle orbital in N-th Landau level is $\psi_m(x,y)=\frac{1}{\sqrt{2^N N! L_y\ell \sqrt{\pi}}} e^{ik_m y} e^{-\frac{(x-k_m\ell^2)^2}{2\ell^2}}H_N(\frac{x-k_m\ell^2}{\ell})$,
where the momentum along the circumference is $k_m=\frac{2\pi m}{L_y}$ and
$m$ labels the orbital center position $x_m=k_m \ell^2$ along the cylinder axis ($\ell=\sqrt{\hbar/eB}$ is the magnetic length).
When the magnetic field is strong, by projecting onto the second Landau level,
the many-body Hamiltonian is written as (see Ref. \cite{supple})
\begin{eqnarray}
	\hat H&=&\sum_{\{m_i\}} V_{m_1,m_2,m_3,m_4}  \hat{a}^{\dagger}_{m_1}\hat{a}^{\dagger}_{m_2}\hat{a}_{m_3}\hat{a}_{m_4}
\end{eqnarray}
where $a^{\dagger}_{m} (a_{m})$ is the creation (annihilation) operator of an electron in the orbital $m$, and
$V$ represents matrix elements of modified Coulomb interaction
$\frac{1}{r} e^{-\frac{r^2}{\xi^2}}$ with a regulated length $\xi=4\ell$ \cite{Zaletel2015}.
Throughout the paper, total filling fraction is set to be  half-filled in the second Landau level
(on top of the fully occupied first Landau level).

\begin{figure}[t]
	\includegraphics[width=0.45\textwidth]{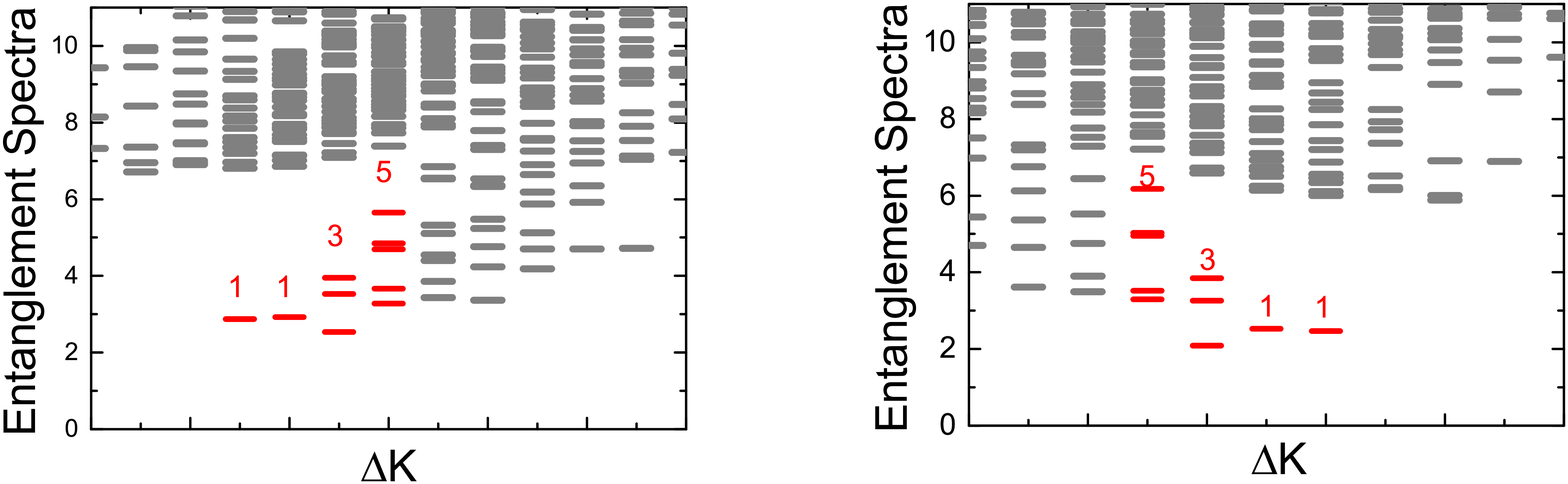}
	\includegraphics[width=0.45\textwidth]{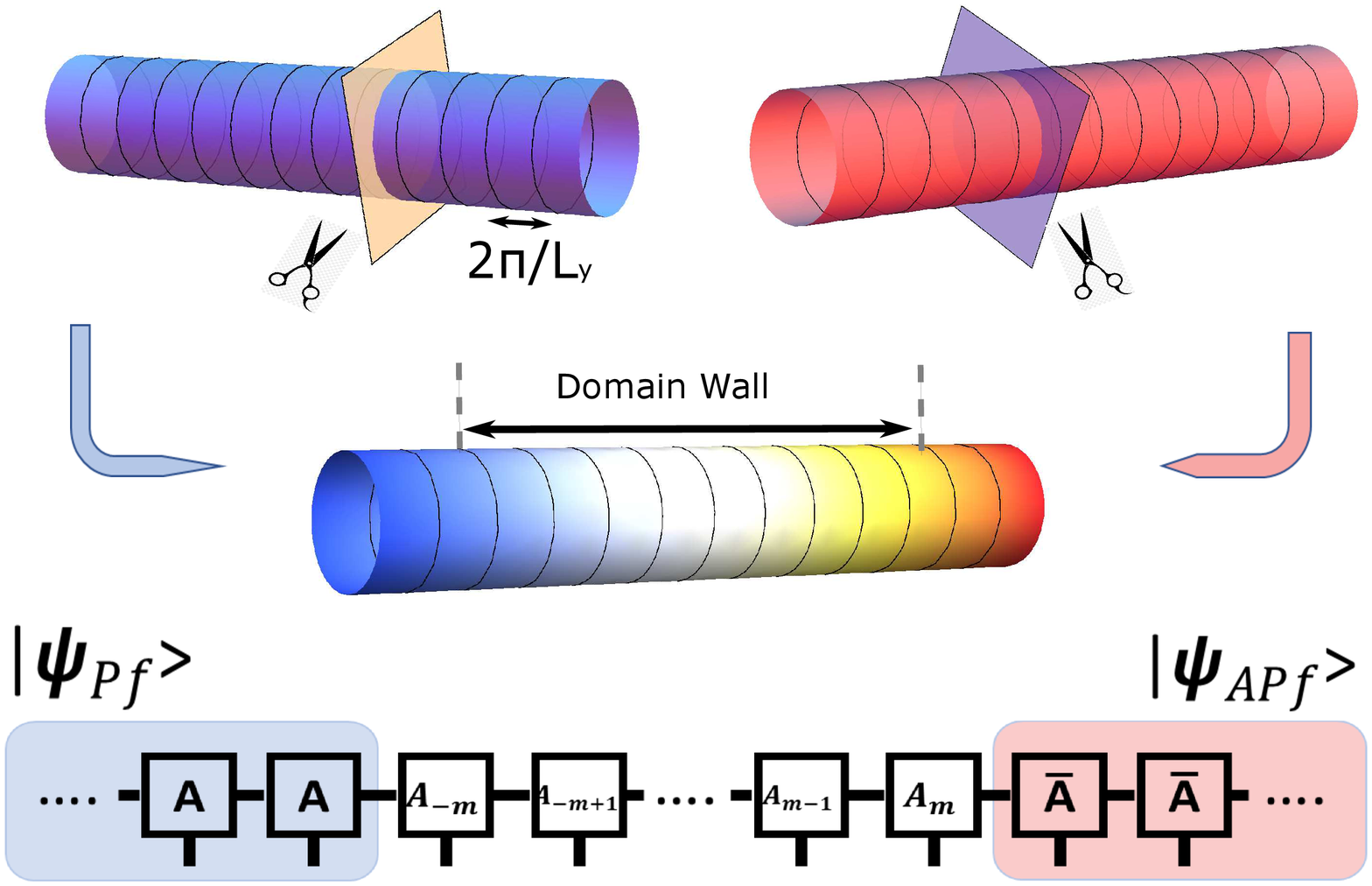}
	\caption{\textbf{Interface between the Pf and APf topological order on the cylinder geometry.}
		(Top) Typical orbital entanglement spectra for gapped Pf (left) and APf (right) state. 
		The chiral dispersions revealed in the entanglement spectra reflect spontaneous breaking of particle-hole symmetry.
		(Middle) Schematic representation of the  Pf-APf interface in the Landau orbital (labeled as black circle) space on the cylinder geometry.
		We first cut the Pf (APf) state into two halves, and glue the left part of Pf state and right part of APf state together,
		which creates an interface regime sandwiched between Pf and APf state.
		(Bottom) The MPS representation of the Pf-APf interface.
	}\label{fig:cylinder}
\end{figure}

For numerical calculations, we  apply a suitable DMRG algorithm with multiple  steps for such an interface system.
The DMRG algorithm is 
based on the matrix product state representation of the ground state:
$|\Psi(\mathrm{A}^{[n_m]}_m) \rangle = ... \mathrm{A}^{[n_0]}_0 \mathrm{A}^{[n_1]}_1...|..., n_0,n_1,...\rangle  $,
where $\mathrm{A}^{[n_m]}_m$ are $D\times D$ matrices and $\{ n_m \}= 0,1$ represents
the occupancy on orbital $m$. In order to model the interface, we  perform the ``cut-and-glue" scheme, %
by  combining finite DMRG \cite{White1992} and infinite DMRG \cite{Ian2008} algorithms, as discussed below.
First, the infinite DMRG is used to iteratively minimize ground state energy
$E_0=\langle \Psi(\mathrm{A}^{[n_m]}_m)|\hat H |\Psi(\mathrm{A}^{[n_m]}_m) \rangle$
by optimizing $\mathrm{A}^{[n_m]}_m$ on an infinite cylinder \cite{Zaletel2013},
which allows us to obtain optimized Pf or  APf state separately. %
The infinite DMRG algorithm has proven to be efficient in the study of FQH ground states
ranging from Abelian to non-Abelian systems \cite{Zaletel2013,Zaletel2015,WZhu2015b}.
Second, based on optimized Pf and APf state living on the infinite cylinder,
we cut both of them into two halves, and then
glue $\mathrm{A}^{[n_m]}_m$ ($m<0$) from the Pf state (shaded in blue)
together with $\mathrm{A}^{[n_m]}_m$ ($m\ge 0$) from the APf state (shaded in red),
which yields an interface between the Pf and APf state
(as graphically shown in Fig. \ref{fig:cylinder}).
Third, by fixing the end of Pf (APf) state as the left (right) boundary,
we optimize the state   on a finite segment enclosing $L_M$ orbitals (up to $L_M=234$)
embedded in the middle of the infinite cylinder.

Here we would like to point out methodological advantages of our scheme.
First, on the infinite cylinder the Pf and APf states are 
automatically selected resulting from spontaneous particle-hole symmetry breaking, and are
treated on equal footing without empirical knowledge. 
Second, one can use established techniques, e.g. entanglement spectra via a cylinder bipartition \cite{Haldane2008,WZhu2016,Zaletel2015},
as a probe of the Pf (APf) topological order (see Fig. \ref{fig:cylinder}(top)).
Third, the microscopic state of the interface can be resolved accurately.
Our calculation is based on a microscopic Hamiltonian
instead of model wave functions \cite{Regnault2019a,Regnault2019b},
so the domain wall structure shown below represents the energetically favorable state at the interface.    

\begin{figure}[b]
	\includegraphics[width=0.45\textwidth]{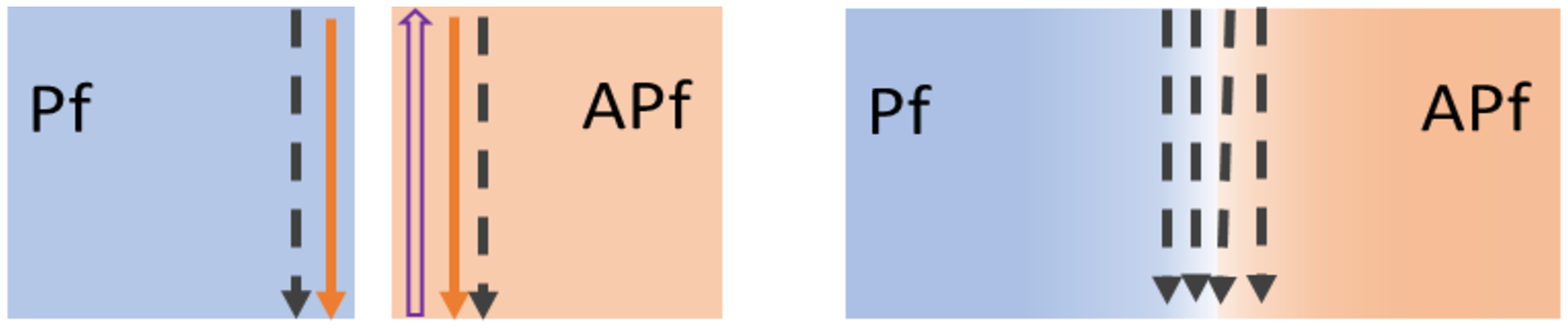}
	\includegraphics[width=0.45\textwidth]{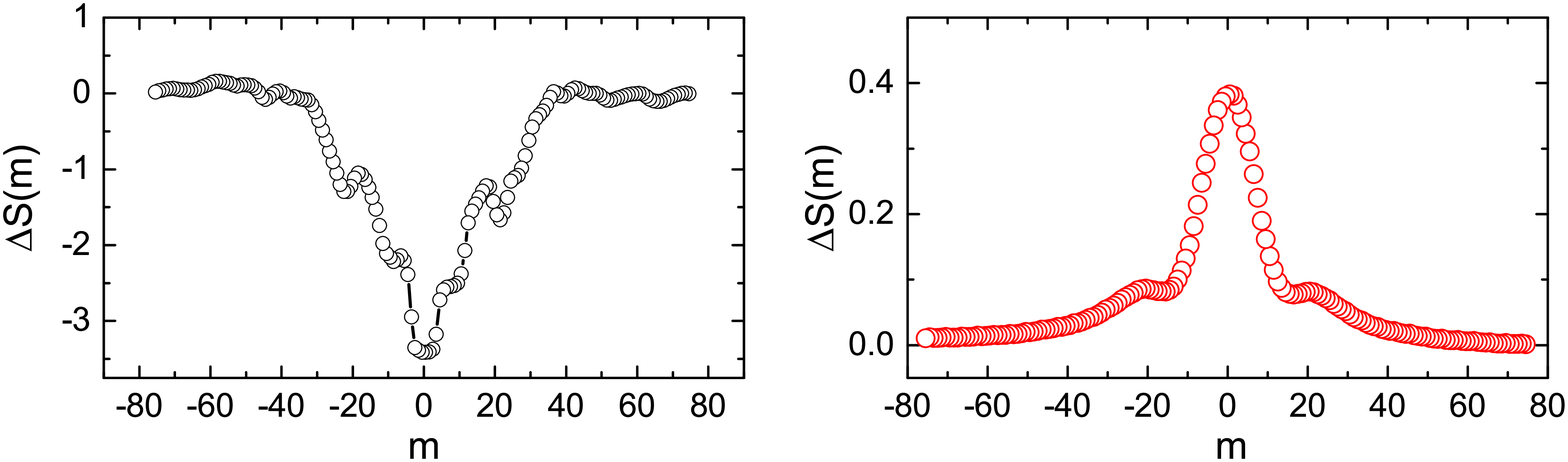}
	\caption{\textbf{Interface structures distinguished by entanglement entropy.}
		The interface structure  in the weak coupling limit (top left)
		and  in the strong hybridization case (top right).
		Here  black dashed lines represent the neutral chiral Majorana fermion modes,
		solid lines represent chiral boson modes of different kinds
		(for details see \cite{supple}).
		(Bottom) The calculated entanglement entropy $\Delta S(m)=S(m)-S_{\textrm{Pf(APf)}}$ dependence on bipartition position $m$.
		($m=0$ is the center of the interface.)
		The entanglement entropy develops a dip in the weak coupling limit (bottom left),
		and a peak structure in the strong hybridization case (bottom right).
		The cylinder perimeter is set to be $L_y=19\ell$ and bond dimension is $D=3600$.
	}\label{fig:edge}
\end{figure}

\textit{Interface structure.---}
We start by discussing the effective edge theories of the Pf and APf state  \cite{Wen1992,Wen1993,Milovanovic1996,Levin2007} (for details see Ref. \cite{supple}).
There are two possible edge structures across the interface depending on the strength of coupling between them \cite{Barkeshli2015}.
If tunneling effect across the interface is irrelevant,
the edge modes of the Pf and APf state form two (nearly) independent sets, sitting on the left and right side of the interface (see Fig. \ref{fig:edge} (top left)).
In this case, if an entanglement measurement is performed, we expect a minimum of the entanglement entropy at the interface, reflecting the effectively decoupled nature between the Pf and APf state.
On the other hand, if the tunneling process across the interface is strong,
the counter-propagating charge modes gap out due to the hybridization effect.
As a result, the Pf-APf interface hosts four co-propagating neutral majorana modes \cite{Mross2018,Chong2018,Lian2018,Wan2016} (see Fig. \ref{fig:edge}(top right)),
allowing neutral fermion to directly tunnel across the interface.
Thus, we expect to see a single smooth peak of entanglement entropy centered at the interface.

Motivated by this intuition, we compute the entanglement entropy and its dependence on the
entanglement cut position, by partitioning the cylinder into two parts at different cut position.
We first create uncoupled edges,  by turning  off   interaction terms acrossing  the interface.
In this case, we observe a dip in entanglement at the interface (Fig. \ref{fig:edge}(bottom left)). 
As a comparison, the result with full  (translationally invariant)  interaction is shown in Fig. \ref{fig:edge}(bottom right).
Far away from the interface, the entanglement entropy converges to the value of the Pf (APf) state.
Near the interface, the entanglement entropy develops a peak centered at the interface.
The appearance of enhanced entanglement across the interface favors
the strong coupling picture (Fig. \ref{fig:edge}(top right)),
and suggests that charged modes are fully gapped out and only neutral modes survive around the Pf-APf interface (see \cite{supple}).

\begin{figure}[t]
	\includegraphics[width=0.425\textwidth]{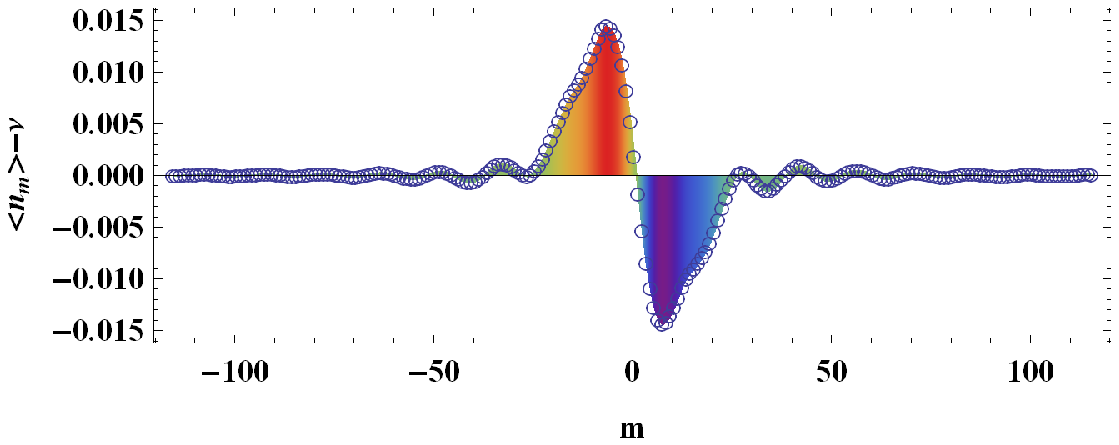}
	\includegraphics[width=0.2\textwidth]{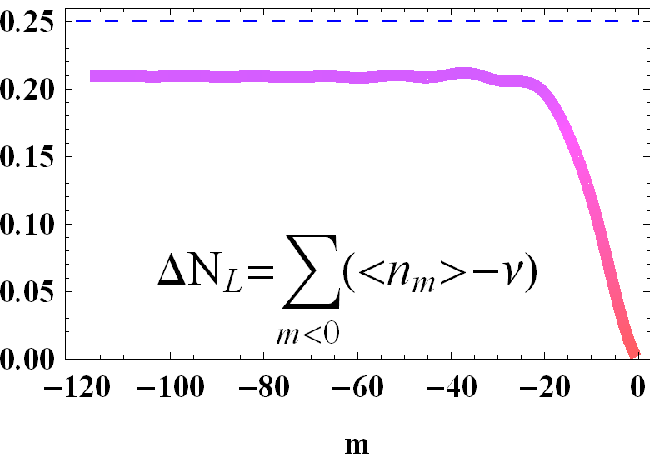}
	\includegraphics[width=0.21\textwidth]{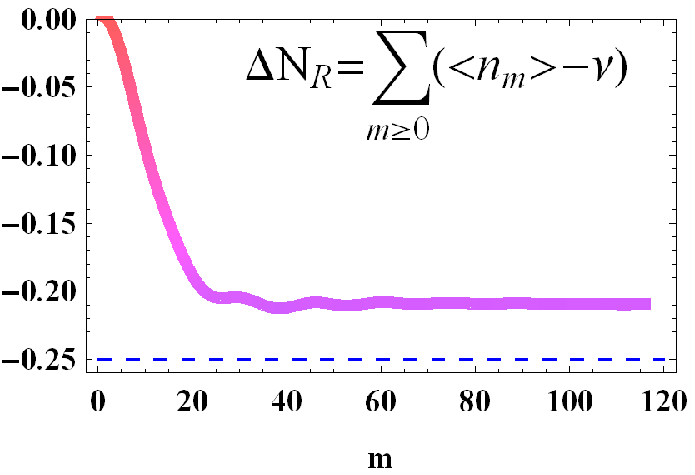}
	\caption{\textbf{Domain wall structure at the interface.}
		(Top) The charge distribution profile  $\langle n_m\rangle -\nu$ (open circles) along the cylinder axis.
		The colored shade denotes the deviation from uniform distribution $\nu=1/2$.
		The leftmost (rightmost) side is a uniform Pf (APf) state.
		(Bottom) The integration of the difference between the actual occupation number and the uniform occupation number
		on the left half $\Delta N_{L}$ (left) and on the right half $\Delta N_{R}$	(right).
		The cylinder perimeter is set to be $L_y=19\ell$ and bond dimension is $D=3600$.
	}\label{fig:Nm}
\end{figure}

\textit{Intrinsic interface dipole moment and Hall viscosity.---}
In spite of the absence of charged modes,
we identify emergent charge fluctuation around the interface.
Fig. \ref{fig:Nm}(top) shows the charge distribution along the cylinder axis.
One salient feature is that charge profile smoothly interpolates between the Pf and the APf state,
and a profound charge fluctuation appears around the interface with small ripples in the periphery of the interface.
In particular, we identify that the Pf (APf) side contains an excess of electron(hole)-like chargers.
(The electron(hole)-rich region switches, if we swap the position of Pf (APf) state in Fig. \ref{fig:cylinder}.)
For quantitative description, Fig. \ref{fig:Nm}(bottom) depicts the accumulation of charge on the left and right side of the interface.
We find several notable features. First, total charge on the left (right) part
of the interface gives  $\Delta N_{L} \approx -\Delta N_{R} $ (in unit of $e$),
where we define the net charge accumulation as 
$\Delta N_{L(R)}  = \sum_{m \lessgtr 0 } [\langle n_m \rangle -\nu]$.
Importantly,  the total charge on each side of the interface is equal but takes opposite sign,
which results in a neutral charge condition without net charge accumulation.
Second, in this particular case the net charge on the left (right) side is very close to 
$\pm e/4$ for a quasi-electron (quasi-hole) as expected for the Pf (APf) state.
Third, we can also identify the domain wall region with a spatial length scale $d_{\textrm{Pf}}+d_{\textrm{APf}}$,
and $d_{\textrm{Pf}} (d_{\textrm{APf}})$ is the distance that domain wall penetrates into the Pf (APf) side.
In Fig. \ref{fig:Nm}(bottom), we estimate $d_{\textrm{Pf}}\approx d_{\textrm{APf}} \sim 8 \ell$.
The obtained spatial penetration depth is slightly larger than previous estimation of quasi-hole radii based on the Pf model wave function \cite{YLWu2014}.
Lastly, we would like to point out, the above finding is similar to that of the ``p-n" junction in semiconductor,
where the neutrality is lost near the p-n interface and the mobile charge carriers form the depletion layer.
Interestingly, different from the p-n junction,
next we will show the origin of charge inhomogeneity at the Pf-APf interface is topological.

We first try to gain some physical intuition of the appearance of electric charge inhomogeneity by considering
 the thin-torus limit \cite{Bergholtz2005,Bernevig2008}.
The typical root configuration pattern of the Pf state is
$...01100110...$, corresponding to a generalized
Pauli principle of no more than two electrons in four consecutive orbitals.
The APf root configuration is simply its particle-hole conjugate. 
In order to switch from one pattern to the other, defects must be introduced near the interface, 
and the simplest one that does not change particle number is
$...0110011_{\times}0|1_{\circ}0011001...$, where
the symbol $\times$ ($\circ$) denotes a quasi-electron (quasi-hole) that
emerges around the nearest four consecutive orbitals
and  $`|'$ labels the interface position.
Therefore, one quasielection-quasihole pair naturally appears around the Pf-APf interface,
providing a direct understanding on the observation of domain wall in Fig. \ref{fig:Nm}.
In addition, since the quasi-electron (quasi-hole) is defined by adding (removing) one electron in
four consecutive orbitals, in the thin-torus limit one can also infer that
the quasielectron (quasihole) carries charge $e^*=e/4$ ($-e/4$). 
The results in Fig. \ref{fig:Nm} largely match that of the thin-tours limit. 
As $L_y$ increases we find the charge transfer increases and deviates from the thin-torus limit, 
however, the dipole moment density of the interface is the intrinsic quantity of topological origin (see below).

The above discussion raises an interesting question: 
Is the formation of a dipole moment intrinsic to the Pf-APf interface? Or, can the quasi-electron and quasi-hole annihilate with each other accidentally?
We now show that the dipole moment at the Pf-APf interface indeed has topological origin by
comparing the topological content of the two bulks.
The Pf (APf) state carries  a different topological number,
the guiding-center Hall viscosity \cite{Avron1995,Haldane2009,Haldane2011,Read2009,Read2011} $\eta^{\textrm{Pf}}_H = -\eta^{\textrm{APf}}_H$,
where the Hall viscosity is determined by the guiding-center spin via $\eta_H= - \frac{\hbar}{4\pi \ell^2}\frac{s}{q}$
(in flat space-time metric).
For the Pf (APf) state, the orbital-averaged guiding center spin takes $\frac{s^{\textrm{Pf}}}{q} = \frac{1}{2}$
and $ \frac{s^{\textrm{APf}}}{q} =-\frac{1}{2}$ \cite{YeJePark2014}, respectively.
Then if the Pf and APf states are put together, there should be a viscous force exerted on a segment of the interface with length $dL_y$:
$dF^{\mathrm{visc}}=(\eta^{\textrm{Pf}}_H-\eta^{\textrm{APf}}_H) B^{-1} \nabla_x E dL_y$
($B$ is the magnetic field and $E(x,y)$ is the non-uniform electric field at the interface).
On the other hand,
around the interface, the electric field coupled with the electric dipole leads to a force:
$dF^{\mathrm{elec}}= \frac{\Delta p^x}{L_y} \nabla_x E dL_y$.
Here, 
we define the dipole moment density as $\Delta p^x/L_y= (p^x(-\infty) - p^x(\infty))/L_y$ and
$\frac{p^x(k)}{L_y} = -e\int^{k}_{0} p\ell^2[\langle n_p \rangle - \nu]\frac{dp}{2\pi}$.
If the interface is stable, we require the above two forces should be balanced $dF^{\mathrm{visc}}+dF^{\mathrm{elec}}=0$.
Therefore, we reach a relationship between the dipole moment density and Hall viscosity:
\begin{align}
\frac{\Delta p^x}{L_y} = B^{-1}(\eta^{\textrm{PF}}_H-\eta^{\textrm{APf}}_H) = - \frac{e}{4\pi} (\frac{s^{\textrm{Pf}}}{q} - \frac{s^{\textrm{APf}}}{q} ) .
\end{align}

\begin{figure}[t]
	\includegraphics[width=0.26\textwidth]{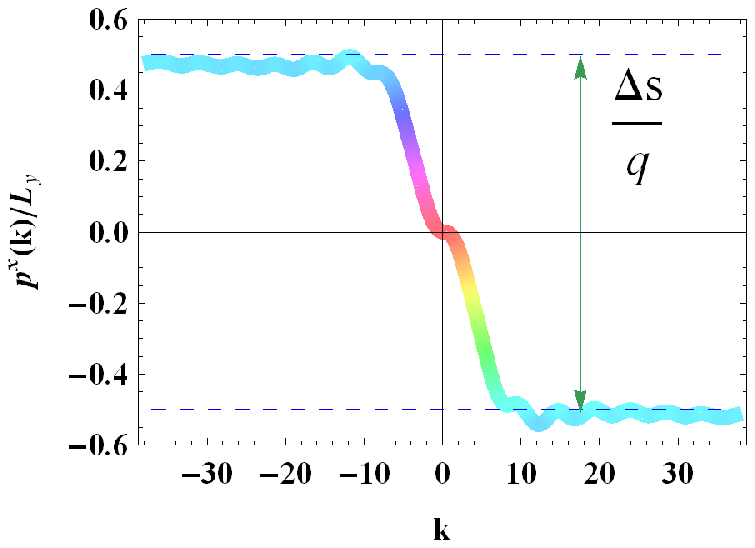}
	\includegraphics[width=0.19\textwidth]{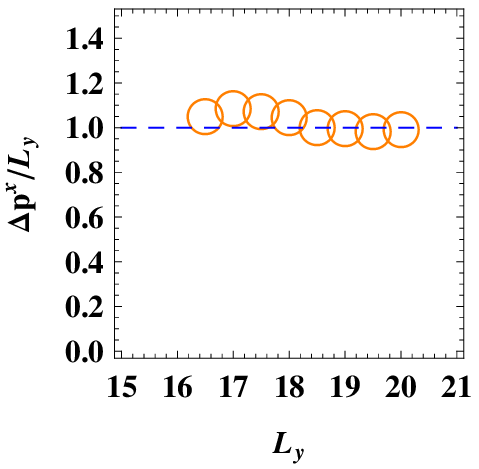}
	\caption{\textbf{Intrinsic dipole moment density near the interface.}
		(Left)	Dipole moment density obtained by
		$p^x(k)/L_y = -e\int^{k}_{0} p\ell^2[\langle n_p \rangle - \nu]\frac{dp}{2\pi}$  (in unit of $(-e/4\pi)$).
		The blue dashed lines show predicted guiding-center spin $\frac{s}{q}=\frac{1}{2} (-\frac{1}{2})$ for the Pf (APf) state.
		The system size is $L_y=19\ell$.
		(Right) The dipole moment density across the interface $\Delta p^x/L_y$ on various cylinder width $L_y$.
		The dashed line denotes the theoretical prediction $\Delta p^x/L_y=\frac{s^{\textrm{Pf}}}{q} - \frac{s^{\textrm{APf}}}{q}=1$.	
	}\label{fig:Mk}
\end{figure}

In Fig. \ref{fig:Mk}(left), we show one typical dipole moment density dependence on momentum $k$ across the interface.
Since the Pf (APf) state is uniform in its bulk, the dipole moment indeed converges to a finite value when $k$ gets large enough.
Crucially, the change of dipole moment density across the interface is $\frac{\Delta p^x}{L_y} \approx 0.99$ (in unit of $(-e/4\pi)$),
close to the guiding-center spin difference $ \frac{s^{\textrm{Pf}}}{q} - \frac{s^{\textrm{APf}}}{q} $. 
In Fig. \ref{fig:Mk}(right), we demonstrate the numerically extracted dipole moment density
for various cylinder width $L_y$, which gets closer to exact quantization with the increase of $L_y$.
As we can see, the dipole moment density is quantitatively in line with theoretical expectation.
Thus our results demonstrate that the formation of electric dipole is to counterbalance
the difference of guiding-center Hall viscosity across the interface.

\textit{Generation of domain wall by disorder.---}
The above discussion demonstrates that the Pf-APf domain wall hosts a intricate structure (see Sec. D.3 \cite{supple}), 
which is overlooked in the effective edge theories \cite{Lian2018,Chong2018,Mross2018,Wan2016}.
It is worth noting that this makes the domain wall  
energetically favorable in the presence of an electric field.
As a result, in real sample sufficiently strong  disorder effect 
could potentially stabilize the Pf-APf domain wall \cite{WZhu2019}.
To be specific, we first estimate the domain wall tension around $\sigma \sim 2\times 10^{-3}e^2/\ell^2$ (Sec. D2 \cite{supple})
(our estimation is largely consistent with a recent work \cite{Simon2019}).
To balance it, the required electric field is around $E_{dis} \ge \sigma/(\Delta p^x/L_y) \sim 2.9\times 10^5 V/m$ (we set $\ell=11.8$nm for $B=5$T).
It is largely in the same order with the typical disorder strength in
the high-mobility GaAs/Ga$_{1-x}$Al$_x$As samples, where local electric field is generated by charged dopants placed about $100$nm from the electron layer. 
Based on this, we conclude the Pf-APf domain wall could be stabilized in the current experimental conditions.

\textit{Summary and discussion.---}
We have presented compelling evidences that
the interface between the Pfaffian (Pf) and anti-Pfaffian (APf) state has intrinsic topological properties.
We identify an inhomogeneous charge distribution around the interface,
where an excess of electron(hole)-like chargers is pinned to the Pf (APf) side,
while the charge neutrality still holds on average.
In particular, the characteristic charge profile yields an electric dipole at the interface,
which is to counterbalance the mismatch in guiding-center Hall viscosity of the Pf and APf state.

Our results unveil a notable effect on the Pf-APf interface, which
is overlooked in the previous discussions \cite{Chong2018,Mross2018,Lian2018,Wan2016}.
This finding may shed lights on the stability of mesoscopic puddles made of Pf and APf order (see \cite{supple}).
In addition, the current work opens up a number of
directions deserving further exploration. For example,
it is an outstanding issue to characterize the topological nature of neutral chiral modes on the interface.
Numerical studies may also further reveal rich physics of the interface made of other exotic non-Abelian states.


\textit{Acknowledgements.---}
W.Z. thanks Bo Yang, Jie Wang, Zhao Liu, Chong Wang, Liangdong Hu for helpful discussion.
W.Z. is supported by project 11974288 from NSFC and the foundation from Westlake University.
D.N.S. was supported by the  U.S. Department of Energy, Office of Basic Energy Sciences under Grant No. DE-FG02-06ER46305.
K.Y.'s work was supported by the National Science Foundation Grant No. DMR-1932796, and performed at the National High Magnetic Field Laboratory, which is supported by National Science Foundation Cooperative Agreement No. DMR-1644779, and the State of Florida.



%


\clearpage
\widetext

\appendix


\beginsupplement

\begin{appendices}

In this supplemental material, we provide more details of the calculation and results to support the discussion in the main text.
In Sec. A, we briefly introduce the effective edge theories that is discussed in the main
text. In Sec. B, we summarize the numerical details of the infinite-size and finite-size density-matrix renormalization
group (DMRG) algorithm on the cylinder geometry. This section includes four subsections.
In Sec. C, we apply the same calculation method on a particle-hole symmetric state,
and show a different picture from the results shown in the main text.
In Sec. D, we discuss the stability of the Pfaffian-anti-Pfaffian domain wall based on our simulations.
This section includes three subsections.
In Sec. E, we present an analysis of edge excitations via entanglement spectra.

\section{A. Effective Edge Theories}
\begin{figure}[b]
	\includegraphics[width=0.25\textwidth]{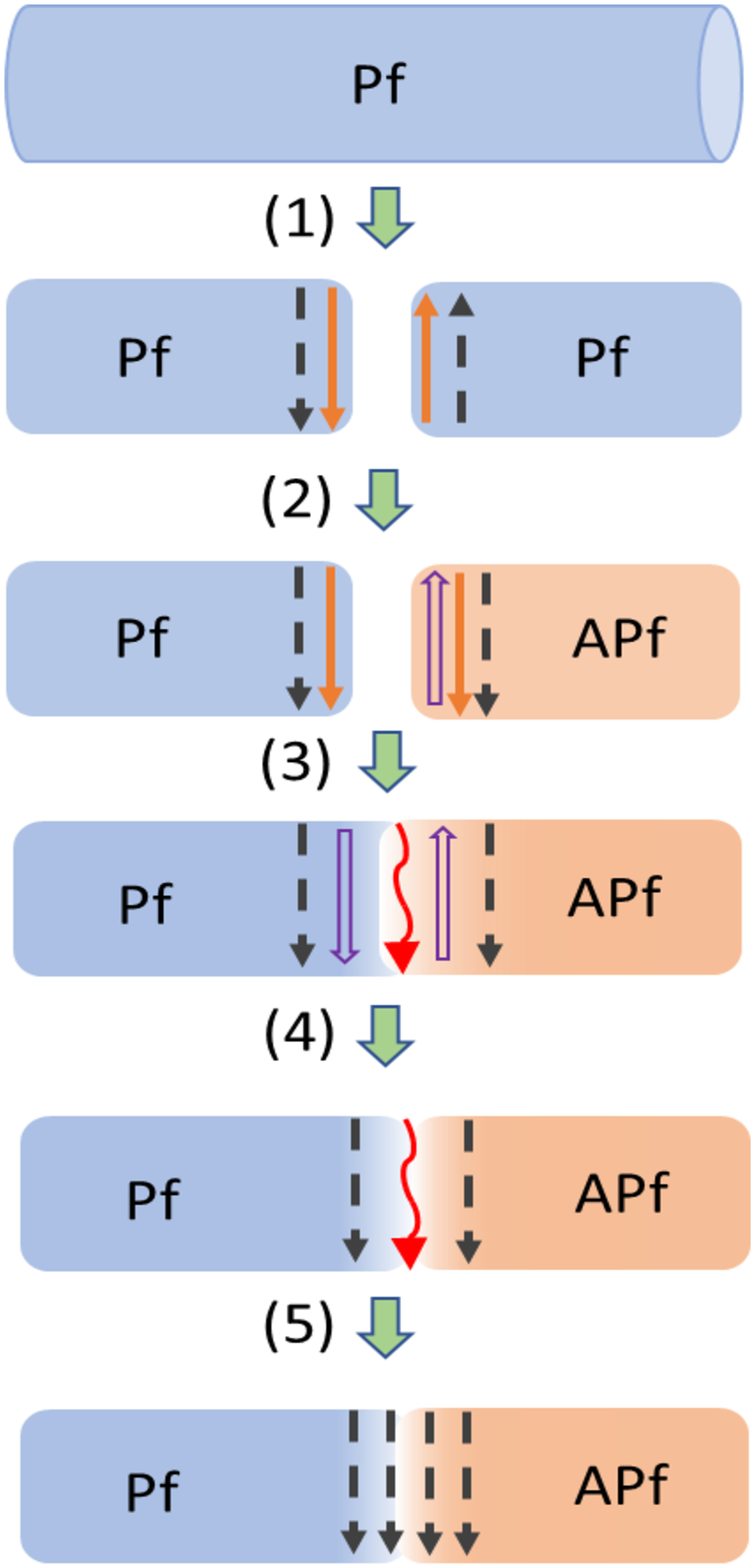}
	\caption{\textbf{Microscopic edge modes at the interface.}
		(1) Starting from the uniform Pf ground state, we make an entanglement cut and bipartition the Pf state into two halves.
		(2) We perform a particle-hole conjugation on the right half, and thus create the Pf-APf interface.
		(3) By introducing the particle tunneling across the interface, the edge reconstruction
		occurs.
		(4) Gapping out the counter-propagating charged modes leads to two co-propagating neutral Majorana modes and one neutral boson mode.
		(5) If further assuming the SO(4) symmetry, the resulting edge theory can be also expressed by four co-propagating neutral Majorana modes.
		Here  black dashed lines represent the neutral chiral Majorana modes,
		solid orange lines represent charge$-e/2$ chiral boson modes,
		and double solid line is a charge$-e$ chiral boson.} \label{sfig:theory}
\end{figure}

In this section, we analyze the effective edge theory of the Pf-APf interface created on the cylinder geometry as shown in the main text (Fig. 2).
As shown in Fig. \ref{sfig:theory}, we first obtain a uniform ground state of the Pf state on an infinite long cylinder.
Then we consider the following process step by step. (The analysis procedure is also shown in Fig. \ref{sfig:theory}.)
\begin{enumerate}
	\item We make an entanglement cut and bipartition the cylinder into two halves.
	At each side, the edge theory is described by $L^{\textrm{left}} = L_{c,+}[\phi_{e/2}] + L_{n,+}[\chi]$ and
	$L^{\textrm{right}} = L_{c,-}[\phi_{e/2}] + L_{n,-}[\chi]$, where $L_{x,\pm}$ describes the charged boson mode ($x=c$) (labeled by orange solid line) and
	the neutral fermion mode ($x=n$) (labeled by black dashed line), and $\pm$ relates to the upstream/downstream mode.
	If we glue the left and right part back, counter-propagating modes are all gapped out and no net edge mode appears near the interface,
	thus recovering the uniform gapped state in the bulk.
	Here the charge boson carries chiral central charge $c=1$, and neutral fermion carries $c=1/2$.
	
	\item We perform a particle-hole operation on the right half of the cylinder,
	and produce a APf state on the right part.
	The particle-hole conjugation demands
	reversing the direction of all edge modes and adding
	another upstream integer edge mode (labeled by purple double solid line):
	$L^{\textrm{right}} =L_{c,-}[\phi_{e}]+ L_{c,+}[\phi_{e/2}] + L_{n,+}[\chi]$.
	If the particle tunnel process is irrelevant, the
	Pf-APf interface hosts two sets of gapless edge modes, and
	they are placed on the left and right sides of the interface, respectively.
	
	\item If we consider the particle tunnel process across the interface,
	the two charge$-e/2$ boson modes would generate a combination, and produce
	a downstream charge$-e$ boson mode  and a boson neutral mode (labeled red wave line):
	$L^{\textrm{Pf-APf}}=L^{\textrm{left}}+L^{\textrm{right}} = L_{c,-}[\phi_{e}] + L_{c,+}[\phi_{e}] + L_{n,+}[\phi_{n}]+ \sum_{i=1,2}L_{n,+}[\chi_i] $
	
	\item  The back-scattering would gapped out two couter-propagating charge modes (double solid lines),
	and leave the neutral modes alone on the interface: $L^{\textrm{Pf-APf}}= L_{n,+}[\phi_{n}]+ \sum_{i=1,2}L_{n,+}[\chi_i] $ .
	
	\item If there is emergent symmetry,
	one can redefine the chiral neutral boson mode as two chiral Majorana
	fermion modes. The neutral Majorana fermions co-propagate and
	thus cannot be gapped out. Consequently, the Pf-APf interface is described by four co-propagating
	Majorana modes：
	 $L^{\textrm{Pf-APf}}=\sum_{i=1,2,3,4} L_{n,+}[\chi_i]$.
\end{enumerate}

The above analysis is in line with the effective theory of Pf-APf stripe state \cite{Wan2016}.
In the main text, we
just present the edge structure before and after the charged modes are gapped out.
The entanglement entropy provides a way to distinguish  if  the tunneling process is irrelevant.

\section{B. Details of the Computational Methods}

In this section, we discuss the details about the numerical simulation.

\subsection{1. Model and Hamiltonian}
We discuss the single electron physics first. In the cylinder geometry,
The coordinate $y$ is along the periodic direction of circumference $L_y$, and $x$ is along cylinder axis direction.
We choose the Landau gauge $\vec{A}=(0,Bx)$ that conserves the momentum around the circumference of cylinder.
In this case, each single electron orbital is labeled by an
integer $m$, with a momenta $k_m$:
\begin{equation}
\psi^N_{m}(x,y)=\left(\frac{1}{2^N N! \pi^{1/2} L_y \ell}\right)^{1/2} \exp[i\frac{X_m}{\ell^2}y-\frac{(X_m-x)^2}{2\ell^2}]H_{N}(\frac{X_m-x}{\ell})
\end{equation}
where $X_m=k_m \ell^2=\frac{2\pi \ell^2}{L_y} m$ is the center along x axis and $\ell$ is the magnetic length.
$H_N(x)$ is the Hermite polynomial and N is Landau level index.

If we project into the second Landau level (setting $N=1$),  the second quantization form of Hamiltonian can be expressed by
\begin{eqnarray} \label{seq:Ham}
	\hat H&=&\sum_{\{m_i\}} V_{m_1,m_2,m_3,m_4}  \hat{a}^{\dagger}_{m_1}\hat{a}^{\dagger}_{m_2}\hat{a}_{m_3}\hat{a}_{m_4},
\end{eqnarray}
where $V_{m_1,m_2,m_3,m_4}$ is the interaction matrix element :
\begin{eqnarray}
&&V_{m_1,...,m_4}=\frac{1}{2L_y} \int^{\infty}_{-\infty} dq_x \sum\limits_{q_y=\frac{2\pi t}{L_y}} F_N(q) V(q_x,q_y)  e^{-\frac{(q^2_x+q^2_y) \ell^2}{2}+iq_x(m_1-m_3)\frac{2\pi \ell^2}{L_y}} \delta_{m_1-m_4,t} \delta_{m_1+m_2,m_3+m_4} .
\end{eqnarray}
$F_N(q)=L_N(\frac{q^2\ell^2}{2})$ is the form factor of N-th Landau level.
The function of $V(q_x,q_y)$ is the Fourier transformation of interaction potential $V(r)$.
In this work, we choose the form of interaction as the modified Coulomb interaction
\begin{equation}
V(|\textbf{r}_1-\textbf{r}_2|) = \frac{1}{|\textbf{r}_1-\textbf{r}_2|} e^{-\frac{(\textbf{r}_1-\textbf{r}_2)^2}{\xi^2}}.
\end{equation}
Here $\xi$ is a regulated length to remove the Coulomb singularity.
It has been carefully checked that, the modified Coulomb interaction can faithfully
capture the essence of physics in fractional quantum Hall systems \cite{Zalatel2015}
(The different choose of $\xi$ doesnot change the physics qualitatively).
In this paper, we will use this modified Coulomb interaction.

\begin{figure}[b]
	\includegraphics[width=0.35\textwidth]{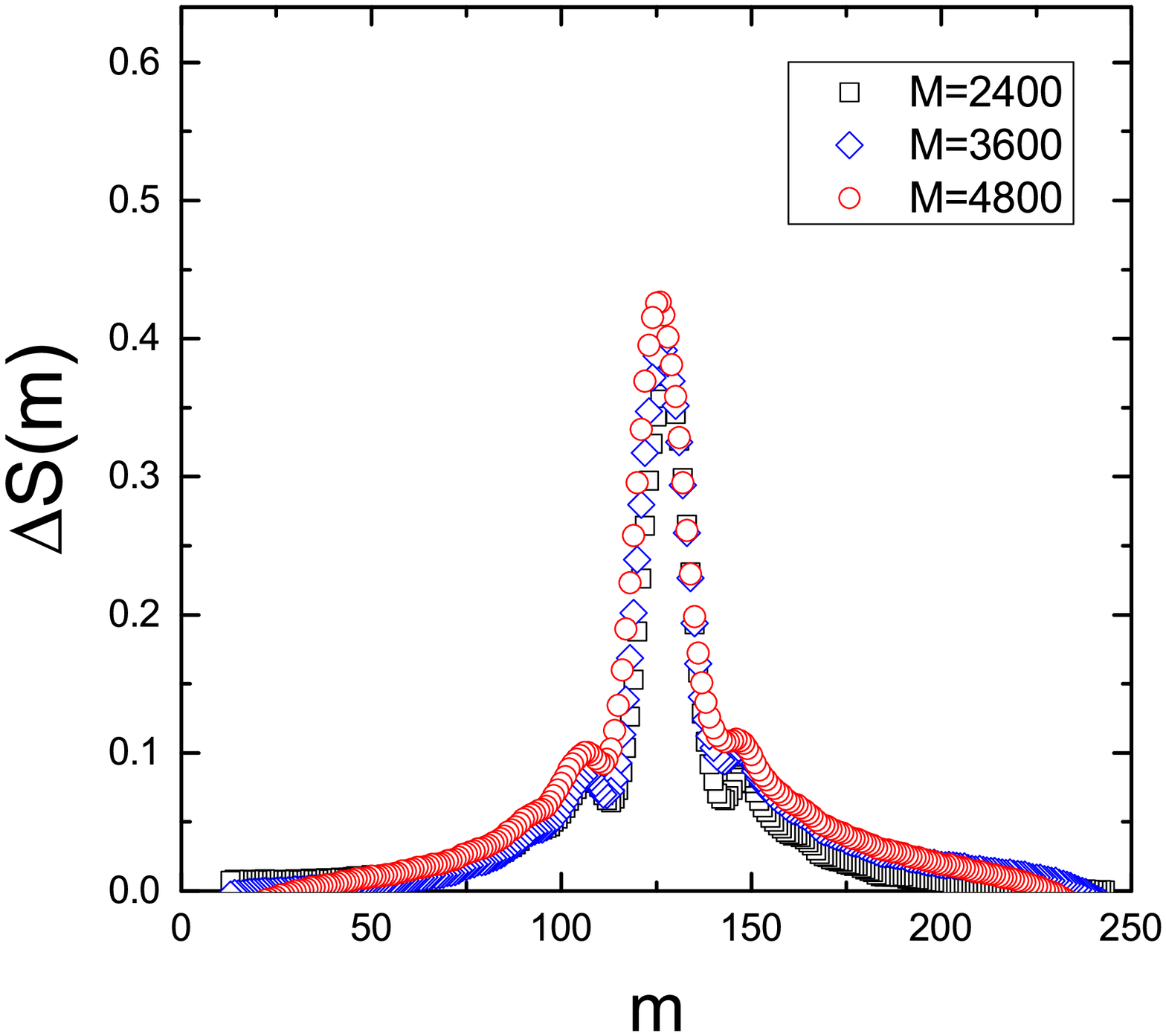}
	\caption{The entanglement entropy $\Delta S= S- S^{\textrm{Pf(APf)}}$ dependence on bipartition position $m$.
		Finite size effect on the domain wall structure obtained by different bond dimensions: $D=2400,3600,4800$.
		Physical measurements show little dependence on the bond dimensions. The system size is set to be $L_y=18\ell$.		
	}\label{sfig:bondM}
\end{figure}

\subsection{2. DMRG calculations}
Previously,
people thought that the cylinder geometry was not suitable for the calculation
of Eq. \ref{seq:Ham}. The reason is, in the traditional DMRG calculation,
to avoid the electrons trapped at the two ends of the finite cylinder,
it is necessary to include an additional one-body potential $U(x)$.
This one-body potential is un-controlled, and its selection is usually empirical.
This issue can be safely overcome by using DMRG on the infinite cylinder geometry \cite{Zaletel2013}.
In infinite DMRG algorithm, one can access the actual results near the center by sweeping,
and the edge effect should be suppressed when the length of the cylinder grows
long enough (infinite long limit). So far, the infinite DMRG has been successfully applied to various FQH states
ranging from Abelian states to non-Abelian states \cite{Zaletel2013,Zaletel2015,WZhu2015b}.

In this work, to study the domain wall between the Pf and APf state,
we combine the finite DMRG and infinite DMRG algorithm.
At the first step, we perform infinite DMRG to get the Pf (APf) ground state.
In all calculations, we do not presume any empirical knowledge from the model wave function.
We reach the same conclusion from a random initial state or an orbital configuration according to the root configuration in the initial DMRG process.
We find that, on the extensive systems with $L_y\in[16,24]$, the DMRG calculation will automatically select
one of the Pf and APf state. 
Once the ground state has been fully developed, we stop the infinite growth of cylinder,
and go to the finite DMRG algorithm.
At the second step,
we glue the Pf state and the APf state together and create a Pf-APf junction (see Fig. 1 in the main text).
We fix the left (right) boundary state as the Pf (APf) state, and perform the finite DMRG variational process in the
central $L_M$ orbitals. $L_M$ changes from $96$ to $234$ in this work to ensure a  converged result for the interface. 

In the implementation, we kept all Coulomb interaction terms $|V_{m_1,m_2,m_3,m_4}|>10^{-6}$ within the truncated range $|m_1-m_2|<4\xi,|m_2-m_3|<\xi L_y/2$.
We have checked that the physical quantities remain qualitatively unchanged
when the truncation range is varied.
In the calculations, we used the bond dimension  kept up to $D=6000$.
We notice that the convergence of the domain wall on $L_y>20$ systems is quite slow,
so that we only present the results with $L_y\le 20$.

\subsection{3. Numerical Identification of the Pf (APf) state}

The  Pf (APf) state can be identified by  its  distinct  edge spectrum. 
Here we analyze the degeneracy pattern of the edge excitation spectrum of the Pf state from the effective edge Hamiltonian.
The edge excitation of the  Pf state contains one branch of free bosons and one branch of Majorana fermions (see Appendix Sec. A),
which can be described by the Hamiltonian \cite{Wen1993}:
$H_{\textrm{edge}}=\sum_{m>0}[E_b(m)b_m^\dagger
b_m+E_f(m-1/2)c_{m-1/2}^\dagger c_{m-1/2}]$,
where $b$ and $b^\dagger$ ($c$ and $c^\dagger$) are standard boson
(fermion) creation and annihilation operators,  and the total
momentum operator is defined as $K=\sum_{m>0}[mb_m^\dagger
b_m+(m-1/2)c_{m-1/2}^\dagger c_{m-1/2}]$.  For even number of fermions, the
edge Hamiltonian leads to a typical edge excitation spectra with counting $1,1,3,5,\cdots$ at momentum point $\Delta K=0,1,2,3,\cdots$.
Here $\Delta K$ is defined as $K-K_0$ where $K_0$ is the lowest momentum ($K_0=0$ for even $F$).

In our calculation, the Pf (APf) state  is identified by the appearance of the typical edge excitation which can be viewed from the entanglement spectra.
As we discussed in Fig. 1 in the main text, the typical entanglement spectra $1,1,3,5,...$ (in the particle counting) gives the evidence of the Pf state,
which relates to the root configuration $...01100110....$. Similarly, the entanglement spectra  $1,1,3,5,...$ (in the hole counting)
signals the APf state. The Pf and APf state has the same counting, but opposite chirality.

In addition, it is known that there are three different topological sectors for the Pf (APf) state: Identity I, neutral fermion f, and Ising anyon $\sigma$.
In this paper, we focus on the identity topological sector (I) which relates to
the root configuration $...01100110....$ with the edge excitation spectrum $1,1,3,5,...$ (as we discussed above).
We construct the interface based on the identity sector of the Pf and APf state.
In principle, one can construct the Pf-APf interface using different topological sectors.
However, the emergent of quasiparticles near the interface may make the interpretation more  complex.
This is out of the current scope, and we will leave it for the future study.

\subsection{4. More Numerical Details}
In the DMRG simulation, the bond dimension parameter $D$ determines the complexity of each matrix product tensor in the calculation,
therefore controls the overall accuracy in the calculations.
The truncation of finite bond dimension is
one source of finite size effects in our computations.
So multiple values of bond dimension and its possible extrapolation to infinity is
a normal scheme to check the physics in the thermodynamic limit for the given cylinder geometry.
In Fig. \ref{sfig:bondM}, we compare the key measurements, entanglement entropy profile around the interface,
for various bond dimensions. It is evident that, the domain wall structure is quite robust,
which is independent of the simulation parameters.
Thus, we reach the conclusion that the observed domain wall structure is intrinsic.

\begin{figure}
	\includegraphics[width=0.325\textwidth]{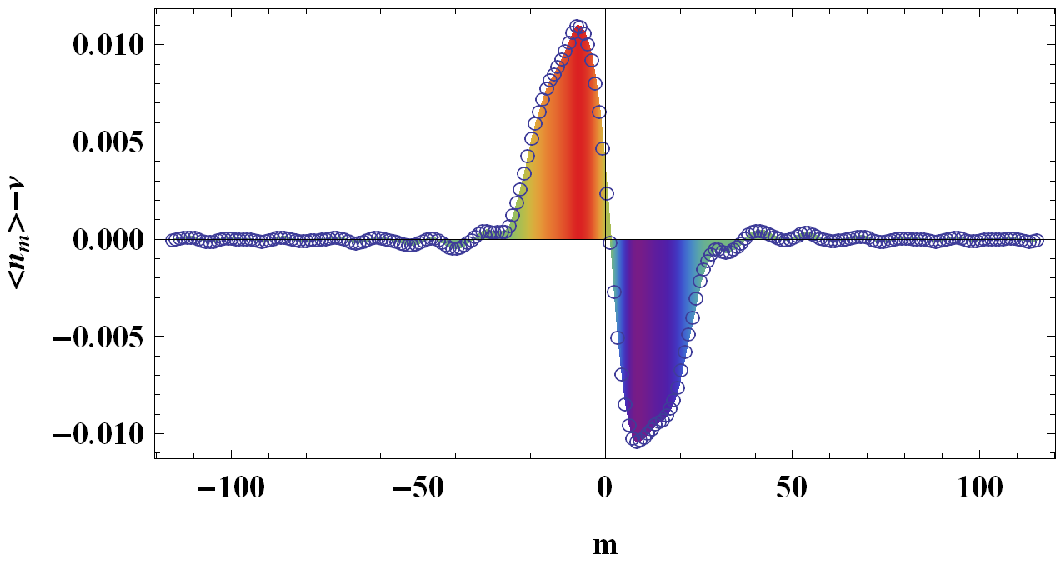}
	\includegraphics[width=0.325\textwidth]{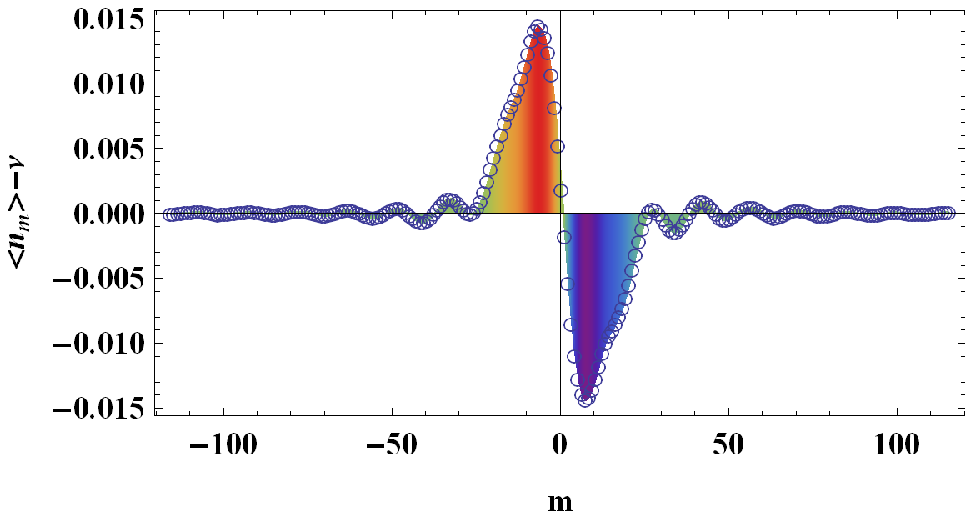}
	\includegraphics[width=0.325\textwidth]{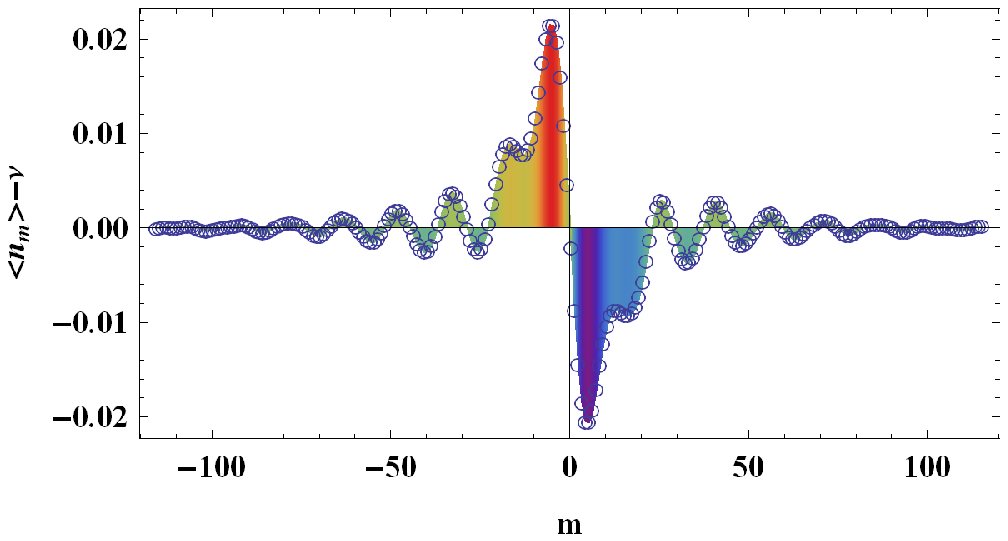}\\
	\includegraphics[width=0.315\textwidth]{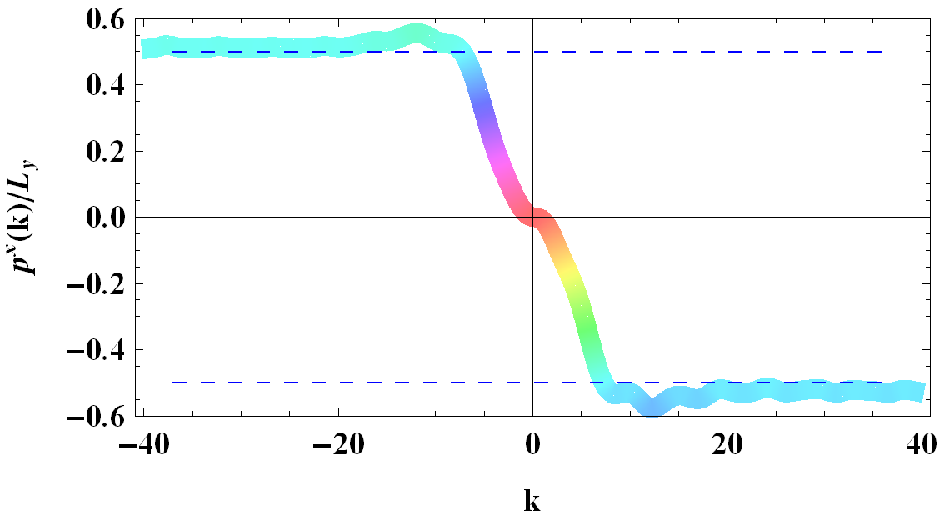}
	\includegraphics[width=0.315\textwidth]{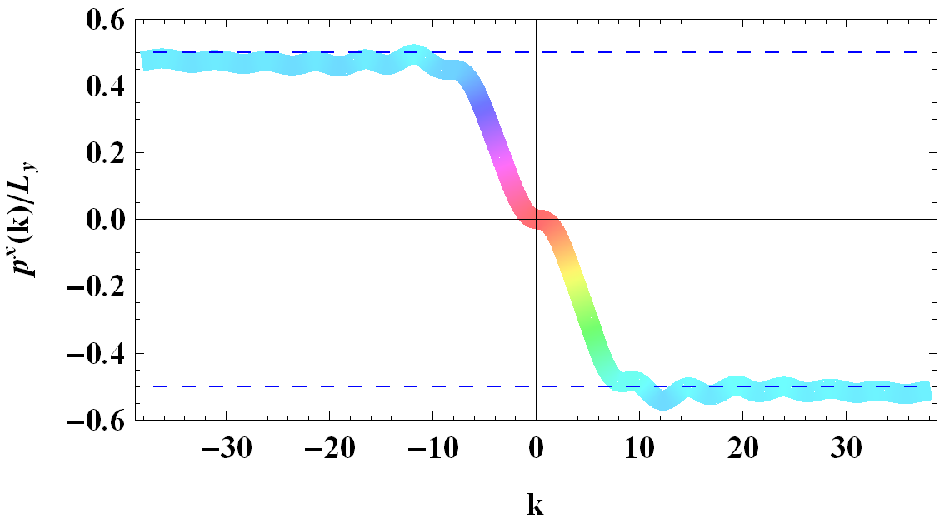}
	\includegraphics[width=0.315\textwidth]{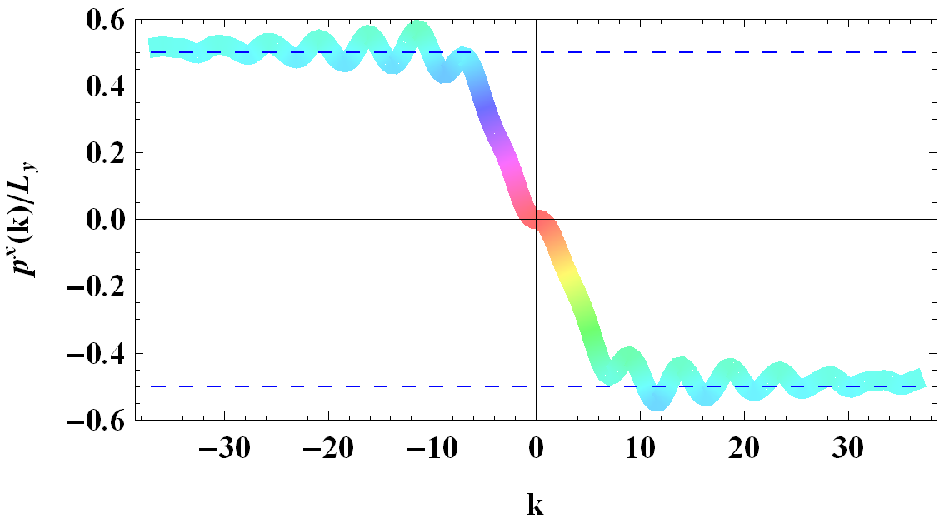}
	\caption{(Top) The charge profile near the Pf-APf interface obtained on various system sizes: $L_y=18$(left), 19(middle), 20 (right).
		$m$ denotes the Landau orbital position. 
		The interface is at the position $m=0$. The leftmost (rightmost) side $m\rightarrow -\infty$ ($m\rightarrow \infty$) is the Pf (APf) state.  
		(Bottom) The dipole moment density obtained by $p^x(k)/L_y $  (in unit of $(-e/4\pi)$) for $L_y=18$(left), 19(middle), 20 (right).
		The momentum is defined by $k = 2\pi m/L_y$.
		The blue dashed lines show predicted guiding-center spin $\frac{s}{q}=\frac{1}{2} (-\frac{1}{2})$ for the Pf (APf) state.
		The bond dimension is set to be $D=2400$.
	}\label{sfig:Ly}
\end{figure}

In addition to the bond dimension, the results on different system sizes
are helpful to infer the physics in the two dimensional limit.   
In Fig. \ref{sfig:Ly}, we show the charge profiles near the Pf-APf interface for
different cylinder width $L_y=18,19,20$.
We see that,a profound domain wall structure can be identified (as we discussed in the main text),
and the domain wall structure largely keeps the similar shape.
In this context, we conclude that the domain wall structure that we reported here, is
quite robust against the finite size effects.

Furthermore, through the comparison in Fig. \ref{sfig:Ly}, we notice that the charge fluctuation becomes larger in the larger system sizes.
For example, on the $L_y=20$ cylinder, it displays ripples in a wider spatial region.
This indicates that, the convergence of the domain wall on larger system sizes is quite slow,
which leads to  much heavier computations on larger systems.
So in this work we only present the results on system sizes $L_y \le 20$.

\begin{figure}[b]
	\includegraphics[width=0.325\textwidth]{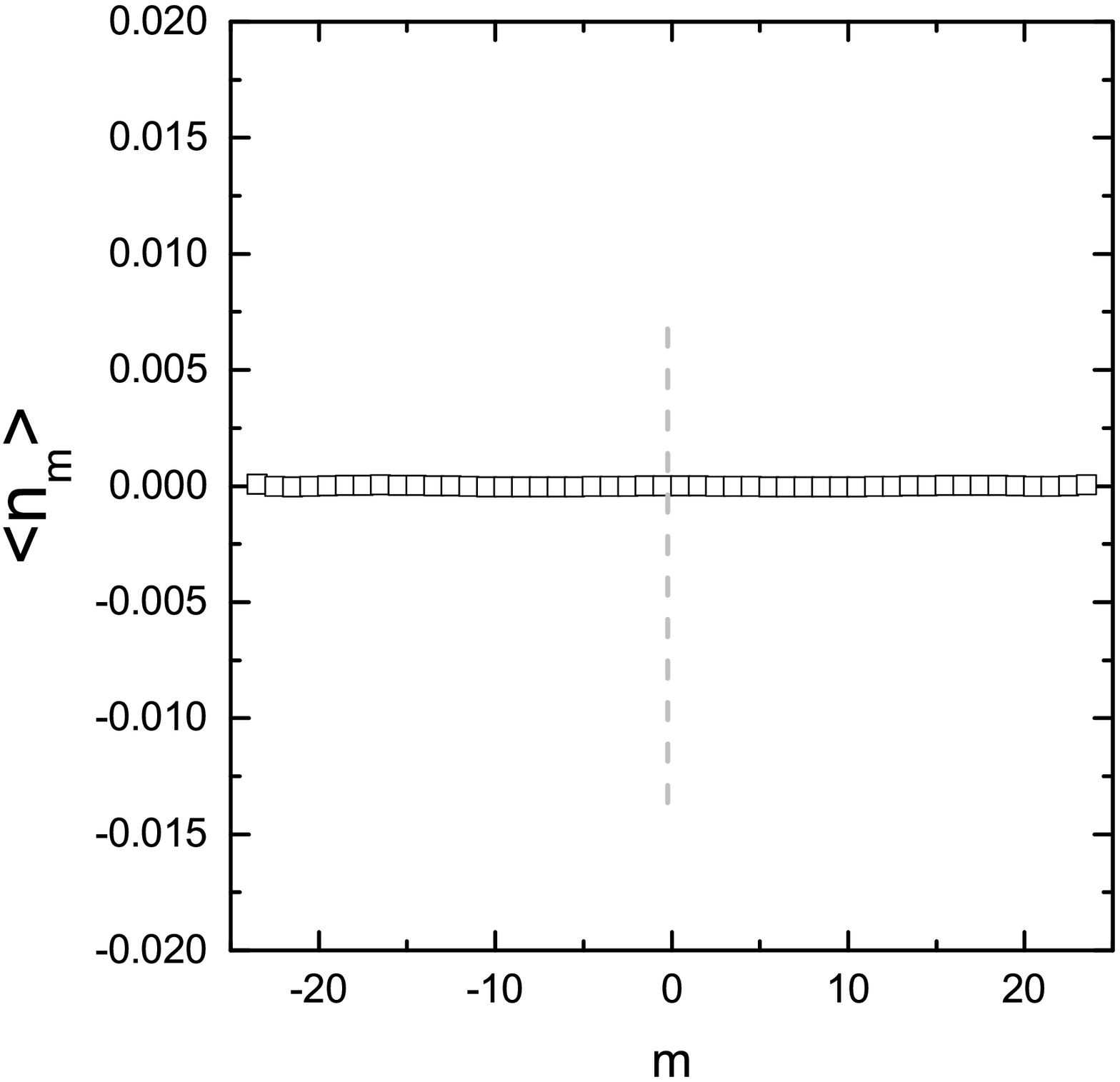}
	\caption{ The charge profile of a particle-hole symmetric state after a particle-hole conjugation is applied on the left half.
		Here we first prepare a particle-hole symmetric state as the ground state first. We
		make a cut and bipartition the ground state into two halves,
		and then apply a particle-hole conjugation operation
		on the left half part (leave the right part unchanged).
		After the DMRG variational process, we measure the charge distribution around the gluing position $m=0$ (marked by the dashed line).
		The particle-hole conjugation process is the same as that in the main text.
	}\label{sfig:density}
\end{figure}

\section{C. Comparison with Particle-hole symmetric state}
In the main text, we elucidate that an electric dipole moment is formed to balance the Hall viscosity difference on the Pf-APf interface.
To further strengthen this point, in this section we study a specific case with no Hall viscosity difference across an interface. As we show below,
no dipole moment forms, if Hall viscosity difference across an interface is zero.

To be specific, we will work on the composite fermion liquid at the half filled $N=0$ Landau level,
which has been proved to be particle-hole symmetric \cite{DTSon2015,Geraedts2016}.
Following the scheme shown in the main text, we make a cut and bipartition the ground state into two halves,
and then apply a particle-hole conjugation operation
on the left half part of the composite fermion liquid, and leave the right part unchanged.
Then we fix the boundary part and make an energy variational calculation in the central part.
The obtained charge profile is shown in Fig. \ref{sfig:density}.
We didnot observe charge inhomogeneity or electric domain wall structure at the gluing position (between orbital $m=-1$ and $m=0$),
which is in sharp contrast to the case discussed in the main text.
The understanding is straight forward: Since the composite fermion liquid is particle-hole symmetric
and its guiding center spin takes $\frac{s}{q}=0$,
no viscosity force is generated thus no electric dipole moment should appear.
In a word, through this test, we further strengthen that,
the formation of electric dipole moment on the Pf-APf interface is intrinsic to the mismatch of the Hall viscosity (guiding-center spin) of the two
distinct topological orders (as we emphasize in the main text).

\section{D. Stability of the Domain Wall: Implications on the random puddles picture}

In this section, we discuss the stability of the Pf-APf domain wall. We address whether or not
it is mechanically or energetically favored in the experimental condition, from the view of numerical simulations.

\subsection{1. Spatial size of the Domain wall}
As discussed in the main text, the domain wall on the interface has a spatial length scale, $d=d_{\textrm{Pf}}+d_{\textrm{APf}}$,
which describes the distance that domain wall penetrates into the Pf (APf) side (see Fig. 3 in the main text).
In our extensive calculation, we estimate this length scale is largely around $d\gtrsim 16\ell$ ($\ell$ is the magnetic length),
depending on the system size $L_y$.
If we take the magnetic length as $\ell = 7.9$nm for external magnetic field strength $B=10$T,
we have the length scale $d \gtrsim 126$nm.
To connect this estimation to the theoretical proposal in Ref. \cite{Chong2018,Lian2018},
where a particle-hole symmetric state is realized by random puddles made of Pf and aPf domains,
we assume each puddle has minimal size $\sim 4d$ ($\sim 2d$ length for boundary and $>2d$ length for separation between two boundary). 
Then we estimate
the minimal size of a puddle made of the Pf and APf state is $\gtrsim 4d \sim 504$nm (for $B=10$T).

\begin{figure}
	\includegraphics[width=0.75\textwidth]{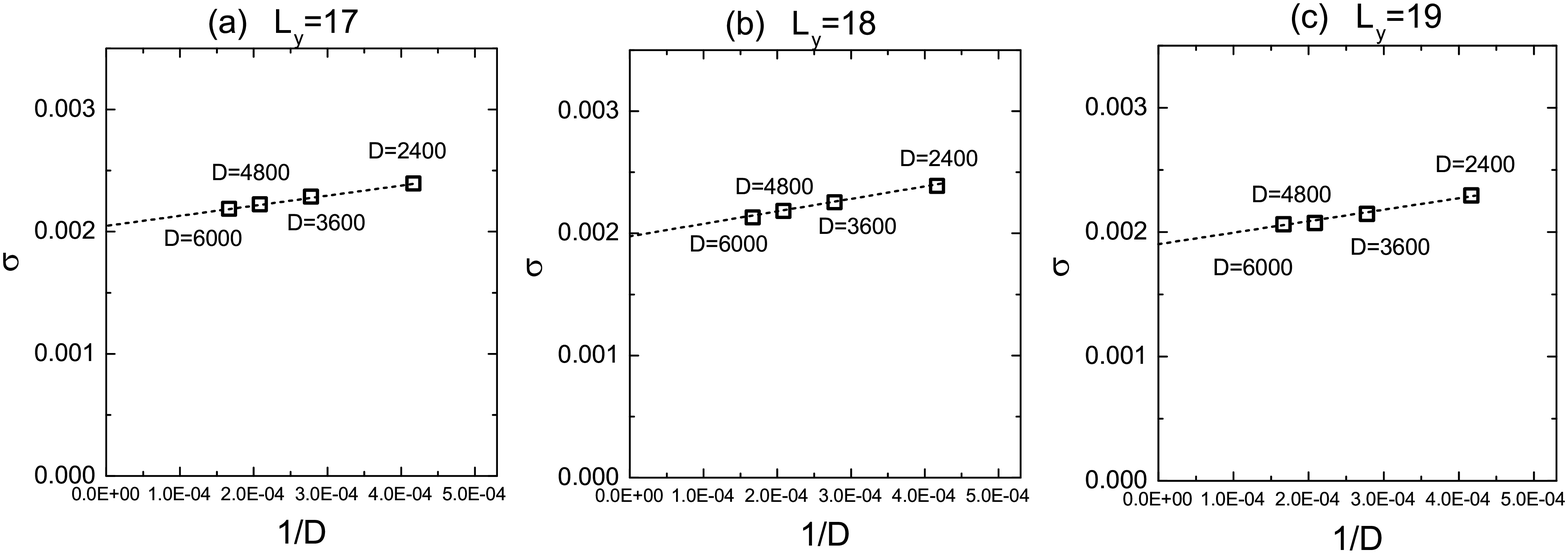}
	\caption{ The tension $\sigma$ (in unit of $e^2/\ell^2$) of a domain wall at the Pf-APf interface,
		obtained on various bond dimension: $D=2400,3600,4800,6000$.
		By extrapolating $\sigma$ using a linear function
		of $D^{-1}$ gives the estimation in $D\rightarrow \infty$.
	}\label{sfig:energy}
\end{figure}

\subsection{2. Energetics of the Domain wall}
In our setup (see discussion in Fig. 1 in the main text), the ground state energy of the Pf (APf) state on the infinite long cylinder is expressed as
\begin{align}
E^{\textrm{Pf}} &= L^{\textrm{Pf}} + C^{\textrm{Pf}} + R^{\textrm{Pf}} ,\\
E^{\textrm{APf}} &= L^{\textrm{APf}} + C^{\textrm{APf}} + R^{\textrm{APf}},
\end{align}
where $L^{\textrm{Pf(APf)}}$ ($R^{\textrm{Pf(APf)}}$) is the energy of the leftmost (rightmost) boundary for the Pf (APf) state,
and $C^{\textrm{Pf(APf)}}$ describes the energy from the central part enclosing $L_M$ orbitals.

Next we consider the Pf-APf domain wall (see Fig. 1 in the main text), sandwiched between a Pf state (on the leftmost side) and
a APf state (on the rightmost side). The obtained energy of this whole system is $E^{\textrm{Pf-APf}}$, which contains
three parts:
\begin{align}
E^{\textrm{Pf-APf}} = L^{\textrm{Pf}} +  C^{\textrm{domain}} + R^{\textrm{APf}}.
\end{align}
Then the energy of domain wall can be derived as
\begin{align}
C^{\textrm{domain}} = E^{\textrm{Pf-APf}} - (L^{\textrm{Pf}} +  R^{\textrm{APf}}) = E^{\textrm{Pf-APf}} - \frac{1}{2} [(E^{\textrm{Pf}} +  E^{\textrm{APf}}) - (C^{\textrm{Pf}} +  C^{\textrm{APf}})].
\end{align}
Therefore, the energy cost of domain wall $\delta=\sigma L_y$ compared to the uniform Pf (APf) state is
\begin{align}
\delta= \sigma \cdot L_y= C^{\textrm{domain}} - \frac{1}{2} (C^{\textrm{Pf}} +  C^{\textrm{APf}})
= E^{\textrm{Pf-APf}} - \frac{1}{2} (E^{\textrm{Pf}} +  E^{\textrm{APf}})
\end{align}
where $\sigma$ is the domain wall tension. 

First of all, in our extensive calculations, the obtained domain wall energy cost $\delta$ are all positive
(in our extensive tests, on all system sizes and calculation parameters,
the domain wall energy costs are positive).
That means, one need to take finite energy cost to (potentially) excite a Pf-APf domain wall structure.
It indicates the formation of Pf-APf domain wall structure is less favored as the ground state (in the translational invariant system),
compared with the Pf or APf state (in a translational invariant system).
(It is further supported by that, we didnot observe any tendency in our DMRG calculation that
the ground state is non-uniform.) 
That is, the other mechanism (e.g. disorder, random potentials) should play some role in stabilizing and favoring
a Pf-APf puddles \cite{Mross2018,Chong2018,Lian2018} (also see discussion below).

In Fig. \ref{sfig:energy}, we compute the domain wall tension $\sigma$,
for a typical cylinder width $L_y=18 \ell, 19\ell, 20\ell$. 
We also extrapolate the calculated results using the $D^{-1}$ ($D$ is bond dimension). 
In Tab. \ref{stab}, we list the obtained domain wall tension on various system sizes.
In a rough estimation, the domain wall tension is around $\sigma \in [0.0017,0.0021]$ (in unit of $e^2/\ell^2$).
The order of this domain wall tension is largely consistent with the recent work \cite{Simon2019}.


\begin{table}[htb] 
	\caption{Estimated domain wall energy and tension on different system sizes.}
	\begin{tabular}{|c|c|c|c|c|c|c|}
		\hline
		\hline
		$L_y (\ell)$ & $17.0$ & $17.5$ & $18.0$ & $18.5$ & $19.0$ & $19.5$ \\ 
		\hline 
		$\delta (e^2/\ell)$  & $0.0349$ & $0.0355$ & $0.0356$ & $0.03381$ & $0.0358$ & $0.0344$ \\
		\hline
		$\sigma (e^2/\ell^2)$ & $0.00205$ & $0.00203$ & $0.00198$ & $0.00183$ & $0.00188$ & $0.00176$ \\
		\hline
	\end{tabular}\label{stab}
\end{table}

\begin{figure}
	\includegraphics[width=0.525\textwidth]{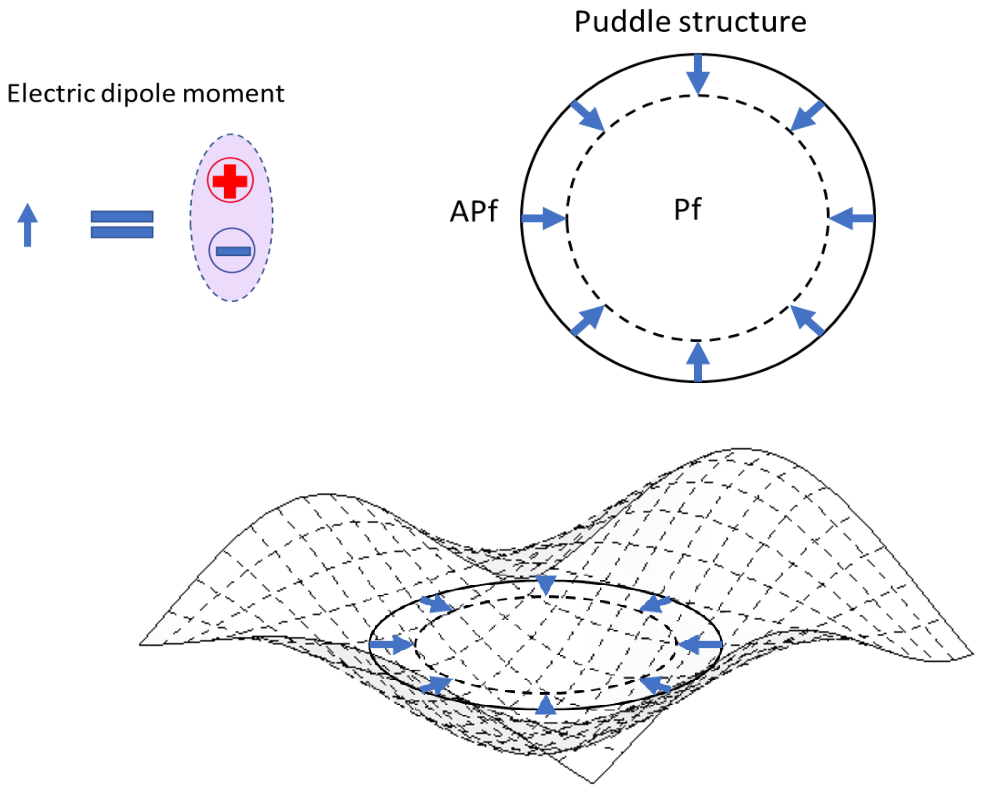}
	\caption{(top) The refined structure of the puddle made of Pf-APf domain wall.
		(bottom) The disorder potential could stabilize the Pf-APf puddle.
	}\label{sfig:puddles}
\end{figure}
\subsection{3. Stability of the Domain-wall}
In Ref. \cite{Chong2018,Lian2018,Mross2018}, it has been proposed a particle-hole symmetric topological order made of domains of Pf and APf state.
Here, our results imply that, if such state is possible, there is a refined structure (Fig.\ref{sfig:puddles}(top left))
which is overlooked in the previous discussion:
The puddle hosts electric dipole moment (denoted by the blue arrow) on the interface (black line) between the Pf and APf state.
The form of this dipole moment has topological origin (as discussed in the main text).
Nevertheless, this puddle structure is not structurally stable,
under the action of a driving force.
For example, considering an external electric field, the coupling between the electric field and
dipole moment requires the dipole moment tends to parallel to the direction of the electric field,
thus the dipole moment structure shown in Fig. \ref{sfig:puddles}(top) is not stable.
Next we argue that, effects of disorder could stabilize the puddle structure.
As shown in the Fig. \ref{sfig:puddles}(bottom),
we assume that disorder creates a relatively weak potential (grey dashed line).
In this case, the puddle structure can be pinned to the equipotential plane of the disorder potential.
The domain wall tension should be at least smaller than the confining potential provided by the disorder potential,
say $\sigma \lesssim (\Delta p^x/L_y) \cdot  E_{dis}$ ($\Delta p^x/L_y$ is the electric dipole density as discussed in the main text).
Thus we estimate that the electric field from disorder potential has the order of $E_{dis}\sim 2.9\times 10^5 V/m$ (we take $\sigma \sim 2\times 10^{-3}e^2/\ell^2$ (see Tab.\ref{stab}), $\ell=11.18$nm at $B=5T$).

At last, we compare this estimated disorder potential with the experimental conditions. 
In the high-mobility $GaAs/Ga_{1-x}Al_xAs$ heterojunction,
the donor layer is usually separated from the two-dimensional electron gas by a typical length scale $d\sim 100nm$.
Thus we estimate the typical disorder potential in experiments as 
$E^{\textrm{exp}}_{dis} \approx k Q/d^2 = (9\times 10^{9} N m^2/C) \times (1.6\times10^{-19}C) /(10^{-7}m)^2 \approx 1.5 \times 10^{-5} V/m $.
Through this estimation, we find that the required disorder potential to stabilize the domain wall is largely 
in the same order with the current experimental condition.

\begin{figure}
	\includegraphics[width=0.725\textwidth]{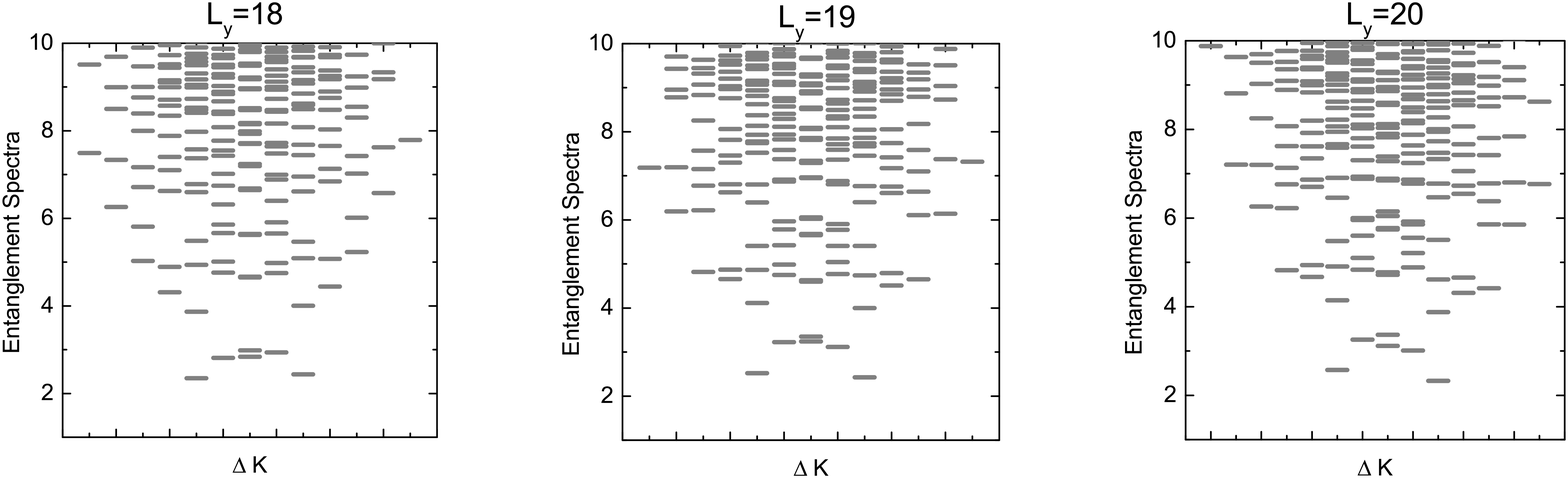}
	\caption{ The entanglement spectra (the entanglement cut position is chosen at the center of Pf-APf interface) obtained on various system sizes: $L_y=18$(left), 19(middle), 20 (right). 	The bond dimension is set to be $D=2400$.
	}\label{sfig:ES}
\end{figure}

\section{E. Entanglement spectra}
In this section, we present the entanglement spectra at the Pf-APf interface.
The typical entanglement spectra at the interface is shown in Fig. \ref{sfig:ES}.
(The entanglement cutting position is at the center of the interface,
where the entanglement entropy reaches a maximum value (as shown in Fig. 2 in the main text).)
Interestingly, it is found the entanglement spectra is almost particle-hole symmetric,
despite of small deviations.
This could be understood, if we recall the root pattern in the thin-torus limit (see the discussion in the main text). 
That is, looking particle excitations from the interface of the Pf side
is similar to looking hole excitations from that of the APf side.

The emergence of particle-hole symmetry at the interface provides a 
numerical self-consistency check of the our computation. 
Importantly, it shows that the center of the interface is special.
Due to this emergence of particle-hole symmetry, the guiding-center spin 
and related guiding-center Hall viscosity at the interface should be zero. 
Thus we can select this point as a reference to compare the guiding-center
viscosity of the Pf or APf state.

\end{appendices}

\end{document}